\documentclass[letterpaper,english,reprint, aps]{revtex4-1}
\usepackage[T1]{fontenc}
\usepackage[latin9]{inputenc}
\usepackage{amsmath}
\usepackage{amssymb}
\usepackage{wasysym}



\usepackage{babel}
\begin{document}

\title{Microscopic origin of de Sitter entropy}

\author{Dmitriy Podolskiy}
\email{Dmitriy_Podolskiy@hms.harvard.edu}

\selectlanguage{english}%

\affiliation{Harvard Medical School, 77 Avenue Louis Pasteur, Boston, MA, 02115}

\date{\today}
\begin{abstract}
It has been argued recently that the entropy of black holes might
be associated with soft scalar, graviton and photon states at the event horizon,
as number of such possible soft states is proportional to the horizon
area. However, the coefficient of proportionality between the number
of soft states and the horizon area of a black hole has not been established.
Here, similar arguments are applied to de Sitter spacetime and it
is shown that soft scalar gravitational modes account for the full
de Sitter entropy $S=\frac{1}{4}M_{P}^{2}A$ with the correct numerical
prefactor in front of the horizon area. We also find how the value
of de Sitter temperature naturally emerges in the treatment of a scalar
quantum field theory on the planar patch of $dS_{4}$. 
\end{abstract}

\keywords{Gravitational entropy, information loss paradox, eternal inflation}
\maketitle

\section{\label{sec:Introduction}Introduction}

Recently, Hawking, Perry and Strominger \citep{Hawking2016} have
argued that the physical origin of black hole entropy \citep{Bekenstein1972,Bekenstein1973,Hawking1975a}
should be associated with soft scalar, graviton and photon supertranslation
hair, carried by a black hole. The associated soft modes are first
excited in the process of gravitational collapse thus carrying information
about particulars of the collapse process. Also, for a formed black
hole, the flux of infalling matter leads to excitation of the soft
modes present on the event horizon, which thus also encode information
about the matter falling into the black hole post collapse phase,
after the event horizon is formed. Presumably, this information is
not entirely lost in the subsequent process of black hole evaporation
being encoded in the phases (and perhaps amplitudes) of outgoing infrared
modes, and the presence of soft supertranslation degrees of freedom
can thus lead to the resolution of the celebrated black hole information
loss paradox \citep{Hawking1976,Mathur2009}. 

Earlier we have also established \citep{Podolsky2012} using the first
quantization picture of gravitational collapse \citep{Vachaspati2006,Greenwood2010}
how exactly such information encoding can happen in the process of
black hole formation: the particle production during the gravitational
collapse leads to a restructuring of the outgoing vacuum state as
observed at spatial infinity; as this particle production is unbounded
in the sense that the occupation numbers for the modes with comoving
frequencies $\omega_{k}<R_{S}$ blow up reaching the asymptotic behavior
$n_{k}\sim\frac{1}{\omega_{k}^{2}}$ at $t\gg R_{S}$, 
and the integral $\int n_{k}d\omega_{k}$ is IR divergent, 
they contribute to the infinite renormalization of the
vacuum state of the asymptotic observer. When one reduces physics
at infinity to observable quantities only, one subtracts this infinite
contribution, but the price paid is the fact that the spatio-temporal
distribution of the phases and amplitudes of the modes with $\omega_{k}<R_{S}$,
which become strongly redshifted at $t\to\infty$ (measured by the
clock of an observer at asymptotic spatial infinity), can be rather
involved. Our practical inability to probe this spatio-temporal structure
is what generates Bekenstein-Hawking entropy: as an observer at spatial
infinity has a very hard time discriminating between vacua with different
IR phase structures, there exists an associated entropy which, as
the authors of \citep{Hawking2016} argued, is proportional to horizon
area. We note that it was impossible to establish the coefficient
proportionality between the entropy and the area using the arguments
of \citep{Hawking2016} alone, and thus it remains unclear whether scalar
degrees of freedom are the only ones which contribute to the Bekenstein-Hawking
entropy or other degrees of freedom (vector, tensor, etc.) can contribute, too. 

Here we would like demonstrate that a logic similar to \citep{Hawking2016},
when applied to the case of de Sitter spacetime, allows one to correctly
estimate its entropy 
\[
S_{{\rm dS}}=\frac{\pi M_{P}^{2}}{H^{2}},
\]
where $H$ is its Hubble scale. As it turns out, de Sitter
entropy is entirely associated with soft gravitational scalar modes,
as counting of the latter gives the correct numerical value for the
ratio of de Sitter entropy and the horizon area of de Sitter spacetime.
In what follows, we shall work in planar patch of 4-dimensional de
Sitter spacetime, although the generalization to higher dimenisions is
straightforward.

\section{\label{sec:LangevinFokkerPlanck}Langevin and Fokker-Planck equations
in planar patch}

As usual, the gravitational perturbation modes in the de Sitter spacetime
can be represented using the ADM split \citep{Arnowitt1960,Mukhanov1992}
\begin{equation}
ds^{2}=-N^{2}dt^{2}+h_{ij}(dx^{i}+N^{i}dt)(dx^{j}+N^{j}dt),
\label{eq:ADM}
\end{equation}
\begin{equation}
h_{ij}=a^{2}[(1+2\zeta)\delta_{ij}+\gamma_{ij}],\,a=a_{0}e^{\int Hdt}
\approx a_{0}e^{H_{0}t}.
\label{eq:ADM2}
\end{equation}
We assume that the only relevant matter degree of freedom is a single
scalar field $\phi$ with the potential $V(\phi)$, so that anisotropic
stress is entirely absent; the perturbations of the scalar field are
denoted by $\delta\phi$. It is well known that in the gauge
\begin{equation}
\delta\phi=0,\,h_{ij}=a^{2}[(1+2\zeta)\delta_{ij}+\gamma_{ij}],\,\partial_{i}\gamma_{ij}=0,\,\gamma_{ii}=0
\label{eq:Gauge1}
\end{equation}
and in the case of exact de Sitter symmetry $H=H_{0}$ the scalar
mode $\zeta$ remains a purely gauge degree of freedom \citep{Maldacena2003}.
In the quasi-de Sitter case $H\ne H_{0},\,|\dot{H}|\ll H^{2}$ it
becomes physical and observable (if de Sitter and post-inflationary
reheating stages \citep{Kofman1997,Greene1997,DavidLyth1998,Podolsky2002a,Bassett2006,Podolsky2006,Dufaux2006}
come to the end, and superhorizon modes of the scalar degree of freedom
$\zeta$ start to reenter the horizon \citep{Mukhanov1992}); this
quasi-de Sitter case is assumed below.

Here, we shall instead consider the gauge $\zeta=0$ such
that the scalar gravitational degrees of freedom are entirely determined
by the perturbations $\delta\phi$ of the scalar field $\phi$. Following
\citep{Starobinsky1986,Starobinsky1994,Enqvist2008}, one can extract
the soft gravitational scalar mode by decomposing the scalar field
as
\begin{equation}
\phi(\log a,{\bf x})=\Phi(\log a,{\bf x})+\int\frac{d^{3}k}{(2\pi)^{3/2}}\theta(k-\epsilon aH)\times
\label{eq:StarobinskyDecomposition}
\end{equation}
\[
\times\left(\hat{a}_{k}\phi_{k}(\log a)e^{-i{\bf kx}}+\hat{a}_{-k}^{\dagger}\phi_{-k}^{*}(\log a)e^{i{\bf kx}}\right)
+\Delta\phi(\log a,{\bf x}),
\]
where $\Phi$ is the soft (superhorizon) component of the scalar field,
the second term in the r.h.s. of (\ref{eq:StarobinskyDecomposition})
determines the subhorizon modes of the scalar field ($\epsilon\lesssim1$
is assumed), and the third term includes corrections suppressed by
powers of the slow roll parameters, which are not accounted for by
the first two terms. The component $\Phi$ can be associated with
the soft scalar gravitational mode, because (a) it is physically impossible
for any observer in a given Hubble patch to discriminate $\Phi$ from
the classical inflaton component and thus from the vaccum contribution,
(b) the inflaton $\phi$ is the only physical scalar gravitational
degree of freedom in the chosen gauge.

It can be shown straightforwardly by substituting the representation
(\ref{eq:StarobinskyDecomposition}) into the equation of motion for
the inflaton that the soft mode $\Phi$ satisfies the Langevin equation
\begin{equation}
\frac{\partial\Phi}{\partial\log a}=-\frac{1}{3H^{2}}\frac{\partial V}{\partial\Phi}+f(\log a),
\label{eq:Langevin}
\end{equation}
where the ``random force'' term $f$ is subject to the relation 
\begin{equation}
\langle f(\log a_{1})f(\log a_{2})\rangle=\frac{H^{2}}{4\pi^{2}}\delta(\log a_{2}-\log a_{1}),
\label{eq:CorrCondition}
\end{equation}
and average is taken over the product of Bunch-Davies vacuum states
$\prod_{k}|0_{k}\rangle$ for the modes $\phi_{k}$ of the scalar
field; the operator $f$ is also quasi-classical in the sense that
it commutes with itself \citep{Starobinsky1986}. Following the usual
prescription, the Fokker-Planck equation 
\begin{equation}
\frac{\partial P}{\partial\log a}=\frac{1}{3\pi M_{P}^{2}}\frac{\partial^{2}}{\partial\Phi^{2}}(V(\Phi)P)+
\label{eq:FokkerPlanck}
\end{equation}
\[
+\frac{M_{P}^{2}}{8\pi}\frac{\partial}{\partial\Phi}\left(\frac{1}{V(\Phi)}\frac{dV}{d\Phi}P\right)
\]
for the probability to measure a given value $\Phi$ in a given Hubble
patch directly follows from the Langevin equation (\ref{eq:Langevin}).
We shall only be interested in the stationary solution of the Fokker-Planck
equation (\ref{eq:FokkerPlanck}) 
\begin{equation}
P(\Phi,\log a\to\infty)=\frac{{\rm Const.}}{V(\Phi)}\exp\left(\frac{3M_{P}^{4}}{8V(\Phi)}\right),
\label{eq:StatSolution}
\end{equation}
which is asymptotically approached as $a\to\infty$; the stationary
solution exists if the integral $\int P(\Phi,\log a\to\infty)d\Phi$
is finite.
\section{\label{sec:Entropy}Calculating entropy}
It is possible to directly extract the de Sitter entropy from the
distribution (\ref{eq:StatSolution}). To do this, consider a space-time
with asymptotic de Sitter symmetries; the de Sitter causal structure (with important corrections discussed below) is realized for 
a spacetime filled with a scalar field with the potential of the form
\begin{equation}
V(\Phi)\approx V_{0}+\delta V(\Phi),\,|\delta V(\Phi)|\ll V_{0},
\label{eq:PotentialPerturb}
\end{equation}
where $V_{0}$ is constant. The Hubble expansion rate is given in
the leading slow roll approximation by 
\begin{equation}
H^{2}=\frac{8\pi}{3M_{P}^{2}}V(\Phi)\approx H_{0}^{2}+\frac{8\pi}{3M_{P}^{2}}\delta V(\Phi),
\label{eq:EinsteinEq}
\end{equation}
and one finds for the exponent in (\ref{eq:StatSolution})
\begin{equation}
\frac{3M_{P}^{4}}{8(V_{0}+\delta V(\Phi))}\approx\frac{3M_{P}^{4}}{8V_{0}}-\frac{3M_{P}^{4}}{8V_{0}}\frac{\delta V(\Phi)}{V_{0}}=
\label{eq:EntropyAndCorrection}
\end{equation}
\[
=\frac{\pi M_{P}^{2}}{H_{0}^{2}}-\frac{3M_{P}^{4}}{8V_{0}}\frac{\delta V(\Phi)}{V_{0}}.
\]
The probability to measure a given value of the background field $\Phi$
in a given Hubble patch is thus given by
\begin{equation}
P(\Phi,\log a\to\infty)\sim\exp\left(\frac{\pi M_{P}^{2}}{H_{0}^{2}}-\frac{3M_{P}^{4}}{8V_{0}}\frac{\delta V(\Phi)}{V_{0}}\right).\label{eq:ProbabilityExpansion}
\end{equation}
The first term in the exponent coincides with the entropy of 4-dimensional
de Sitter space. Indeed, for the $dS_{4}$ entropy one has $S=\frac{M_{P}^{2}A}{4}$,
where $A=\frac{4\pi}{H_{0}^{2}}$ is the horizon area of de Sitter
spacetime with the Hubble expansion rate $H_{0}$. We see that the
standard prefactor $1/4$ for the gravitational entropy is recovered
correctly. Then, expanding in small $\delta{}V$ one can also write
\begin{equation}
P(\Phi,\log a\to\infty)\sim{\rm Const.}\exp\left(S-\frac{8\pi^{2}\delta V(\Phi)}{3H_{0}^{4}}\right).
\label{eq:ApproxDistrib1}
\end{equation}
Let us take a closer look at the second term in the exponent in
(\ref{eq:ApproxDistrib1}). It is well known that $T=\frac{H_{0}}{2\pi}$
is the temperature of radiation as experienced by the observer in
the static patch of de Sitter spacetime \citep{Spradlin2001}. Thus,
it is convenient to rewrite the second term in the exponent (\ref{eq:ApproxDistrib1})
as $\frac{4\pi\delta V}{3H_{0}^{3}T}.$ The factor $\frac{4\pi}{3H_{0}^{3}}$
in turn coincides with a volume of the 3-dimensional sphere with the
radius $H_{0}^{-1}$, i.e., the comoving 3-volume cross-section $v_{{\rm dS}}$
of a single Hubble patch. We finally conclude that 
\begin{equation}
P(\Phi)\sim{\rm Const.}\exp\left(S_{{\rm dS}}-v_{{\rm dS}}\frac{\delta V(\Phi)}{T}\right)=
\label{eq:ApproxDistrib2}
\end{equation}
\[
={\rm Const.}\exp\left(S_{{\rm dS}}-\frac{F(\Phi)}{T}\right),
\]
where the $F=v_{{\rm dS}}\delta V(\phi)$ is the ``free energy''
of the scalar field slowly rolling towards the minimum of its effective
potential as perceived by a physical observer living within the given
Hubble patch.

We note in passing that the expression similar to (\ref{eq:ApproxDistrib1})
can be derived for arbitrary matter content introduced in the de
Sitter universe given that the energy density of the added degrees
of freedom coarse-grained at the Hubble scale is sufficiently small ($\epsilon\ll V_{0}$);
such coarse-grained energy density should again be understood as a
correction to the vacuum energy density as perceived by an observer
living within a given Hubble patch. In that case, the ``free energy''
will acquire an additive correction $\sim\epsilon\cdot v_{{\rm dS}}$.

For completeness, we shall also derive here the expression for de
Sitter entropy in an arbitrary number of dimensions. Using stochastic
formalism for a $D$-dimensional quasi-de Sitter spacetime \citep{Podolsky2010},
we find for the asymptotic probability to measure a given value of
the inflaton field $\Phi$ in a given Hubble patch
\[
P(\Phi,\log a\to\infty)\sim V^{-1}\exp\left(\frac{\pi^{\frac{D-1}{2}}}{\Gamma\left(\frac{D-1}{2}\right)}\left(\frac{M_{P}}{H}\right)^{D-2}\right).
\]
Applying the same arguments as presented above for the case of $4$-dimensional
de Sitter spacetime and recalling the expressions for the volume of
a $d$-dimensional ball as well as the area of a sphere serving as
its boundary, we also conclude that (a) de Sitter entropy is universally
given by
\begin{equation}
S_{{\rm dS}}=\frac{1}{4}M_{P}^{D-2}A\label{eq:DdimEntropy}
\end{equation}
with the same numerical prefactor $1/4$ in all dimensions as should
have been expected, while the de Sitter temperature is
\begin{equation}
T_{{\rm dS}}=\frac{(D-2)H_{0}}{4\pi}.\label{eq:DdimTemperature}
\end{equation}
\section{De Sitter entropy in the presence of disorder\label{sec:RandomPotential}}
Before proceeding to the discussion of the physical implications of
the formulae (\ref{eq:ApproxDistrib1}) and (\ref{eq:ApproxDistrib2}),
let us consider how the expression (\ref{eq:ApproxDistrib2}) changes
in the presence of disorder in the potential $V(\Phi)$. The main
motivation to ask this question is the fact that within the framework
of string theory one expects to find a very large landscape of de
Sitter vacua for moduli and other scalars able to support eternal
inflation \cite{Bousso2000,Susskind2003,Douglas2003,Denef2004,Garriga2005,Podolsky2007a,Podolsky2008a,Podolsky2009a}.

Considering a particular Hubble patch (that is, focusing on physical
questions asked by an observer living in it), it is also natural to
expect that in the limit $\log a\to\infty$ most such vacua will be
visited, and the resulting potential of the effective inflaton field
will be strongly disordered. (Here by the inflaton one can understand a master scalar field, obtained
from the effective potential of scalars present in string theory by
model reduction \citep{Antoulas2009}.) Moreover, it will be possible to obtain the effective probability
$P(\Phi)$ to measure a given value of the effective inflaton field
$\Phi$ by averaging over disorder (exactly because in the limit $\log a\to\infty$
most quasi-de Sitter vacua on the landscape are already visited). 

Consider for simplicity a random potential of the form
\[
V(\Phi)\approx V_{0}+\delta V(\Phi),\,\,|\delta V(\Phi)|\ll V_{0}
\]
distributed according to a Gaussian measure
\begin{equation}
\int{\cal D}\delta V\exp\left(-\int dx\sqrt{g}\frac{\delta V^{2}}{2\tilde{\Delta}}\right)\sim
\label{eq:GaussianDisorder}
\end{equation}
\[
\int d\delta V\exp\left(-v_{dS}\frac{\delta V^{2}}{2\tilde{\Delta}}\right)=\int d\delta V\exp\left(-\frac{\delta V^{2}}{2\Delta}\right),
\]
where integration in the first expression is taken over the spatial
cross-section (comoving 3-volume) of the planar patch, and $\Delta=\frac{\tilde{\Delta}}{v_{{\rm dS}}}=\frac{3\tilde{\Delta}H_{0}^{3}}{4\pi}$.
(The distribution function for disorder will be of course of a generic
non-Gaussian form, but non-Gaussian corrections to the integral (\ref{eq:GaussianDisorder})
are expected to be suppressed in the ultralocal approximation effectively
realized for a given observer in the quasi-de Sitter universe.) 

As we already mentioned, from the point of view of an observer in
a given Hubble patch the effective value of the cosmological constant
$\sim V_{0}$ is determined by the fact that the field $\Phi$ visits
all possible configurations of disorder realized on the landscape.
We thus find
\begin{equation}
\langle P(\Phi)\rangle_{{\rm disorder}}\sim\int d\delta V\frac{1}{V_{0}+\delta V}\times
\label{eq:AveragingOverDisorder}
\end{equation}
\[
\times\exp\left(\frac{3M_{P}^{4}}{8(V_{0}+\delta V)}\right)\exp\left(-\frac{\delta V^{2}}{2\Delta}\right).
\]
The integral in (\ref{eq:AveragingOverDisorder}) can be calculated
using the saddle point approximation, and one finds that 
\begin{equation}
\langle P(\Phi)\rangle_{{\rm disorder}}\sim\exp\left(\frac{3M_{P}^{4}}{8V_{0}\left(1-\frac{3M_{P}^{4}\Delta}{8V_{0}^{3}}\right)}\right),\label{eq:ResultDisorder}
\end{equation}
i.e., a weak disorder present in the scalar field potential leads to a slight increase in de Sitter entropy,
which is of the order $\delta S_{{\rm dS}}\approx S_{{\rm dS}}^{2}\frac{\Delta}{V_{0}^{2}},$
where $\frac{\Delta}{V_{0}^{2}}S_{{\rm dS}}\ll1$ and $S_{{\rm dS}}=\frac{3M_{P}^{4}}{8V_{0}}$
\textemdash{} the expression (\ref{eq:ResultDisorder}) holds as long
as disorder is sufficiently weak (we naturally expect $S_{{\rm dS}}\gg1$).
\section{\label{sec:Conclusion}Discussion}
We have seen above that the stochastic formalism describing IR dynamics
of a scalar field in quasi-de Sitter universe can serve as a window into
gravitational de Sitter thermodynamics. In particular, it allows
to calculate de Sitter entropy with the correct numerical prefactor
and obtain a correct value for the temperature of de Sitter radiation
despite the fact that the spectra of scalar field modes in the planar
patch of de Sitter spacetime are manifestly non-thermal \citep{Mukhanov1992}.
It also makes it possible to finally address an old question how the
thermal physics of the static patch (a string theorist's favorite) of de Sitter spacetime \citep{Gibbons1977}
is related to the physics of planar patch (a cosmologist's favorite) \citep{Frolov2003}. This
question is important as the actual causal structure of self-reproducing
quasi-de Sitter inflationary universe is very different from the one of static patch:
one of the diamonds is entirely absent from the resulting Penrose diagram (and thus tracing out the corresponding
degrees of freedom in order to obtain thermal flux is less than straightforward), while the other
is modified into an infinite self-replicating set. Where does the ``thermal flux'' and
de Sitter entropy then come from and is it even physical? The correct
answer is that it is physical, and the de Sitter entropy comes from
effective averaging over superhorizon modes constantly generated
in the quasi-de Sitter universe (i.e., the Fokker-Planck distribution
(\ref{eq:FokkerPlanck})): if accelerated expansion never comes
to the end, a physical observer living in a given Hubble patch is
unable to discriminate between the true vacuum state (the direct product
of Fock vacuum states for individual modes) and the state containing
arbitrary configurations of superhorizon modes. Counting these effectively
degenerate states produces correct value for de Sitter entropy, and
we would like to identify such states as a de Sitter analogue of Hawking-Perry-Strominger
soft hair for black hole. The degeneracy is quasi-classical, as it is generated
by the Langevin equation (\ref{eq:Langevin}), only containing terms
commuting with each other and having zero self-commutators at coincident
points (note although that when deriving (\ref{eq:Langevin}) we coarse-grain
spacetime at the Hubble scale, physically rather significant; in this
sense, UV point splitting is a splitting between two different causally
unconnected Hubble patches). This is a direct consequence of the mode
freeze-out after the horizon crossing \citep{Mukhanov1992}. Several
other comments on the physical consequences of this result are in
order:

1) \textbf{``What does de Sitter spacetime evaporate into?''} If
a black hole emitting thermal flux from the event horizon inevitably
evaporates \citep{Hawking1976}, and the physical picture of the static
patch of de Sitter spacetime is also the one supporting a thermal
flux from the horizon \citep{Gibbons1977}, what is the final outcome
of thermal evaporation process in this case? If $V_{0}>0$, the answer
is eternally inflating universe with self-reproducing causal structure;
the pure de Sitter spacetime is indeed unstable \citep{Polyakov2012}, but
the physical origin of this instability is with respect to changing
the matter content of the theory (or even introducing perturbations
to $V(\Phi)$, so that $V(\Phi)\ne V_{0}$): theories with a slightly different matter content will lead
to a widely different self-replicating causal structure of spacetime
at $\log a\to\infty$. 

2) \textbf{Is the physical picture of gravitational thermodynamics
universal? Is it thermal picture applicable to the global causal structure of the
resulting spacetime?} While the common opinion is that it absolutely
is \citep{Jacobson1995,Jacobson2016,Bousso2002,Verlinde2000,VanRaamsdonk2009,Padmanabhan2009,Lashkari2013}, 
we would like to take an opposite point of view motivated by two facts. Namely, (a) if $V_{0}=0$ and
inflation comes to the end being replaced by a decelerating FRW expansion,
the Fokker-Planck equation (\ref{eq:FokkerPlanck}) does not admit
a normalizable time-independent asymptotic solution of the form (\ref{eq:ProbabilityExpansion});
instead, the general solution of the Fokker-Planck equation decays
to zero as $\log a\to\infty$. As such, there is no thermal flux and
no associated entropy hard-wired into the theory, which is in a sense
natural: once inflation ends and is replaced by a decelerated expansion
continued ad infinitum, all inflationary modes eventually reenter the cosmological
horizon, and the whole inflationary history can be potentially recovered
by a sufficiently long-living and patient observer. Naturally, it
is impossible to argue for the thermal equilibrium (and thermal flux)
when the characteristic temperature behaves non-adiabatically, $\dot{T}\apprge T^{2}$
(such is the case for a decelerating FRW universe). Also, (b) the
``Hawking-Moss instanton''-like distribution (\ref{eq:ProbabilityExpansion})
is manifestly non-Boltzmann \citep{Hawking1981}, and while the leading terms in $\frac{\delta V}{V_{0}}$
(or $\frac{\delta\epsilon}{V_{0}}$) expansion do correspond to a
quasi-Boltzmann form (\ref{eq:ProbabilityExpansion}),
higher order expansion terms spoil it. 

3) \textbf{Existence of a non-trivial gravitational entropy is determined
by accessibility of information.} As we just argued, information encoded
in amplitudes and phases of inflationary modes can become accessible
if inflation comes to the end, universe passes through reheating stage
and inflationary modes which reenter the cosmological horizon become
available to probe. A counterpart of the Fokker-Planck equation (\ref{eq:FokkerPlanck})
for cosmology including a decelerating FRW branch does not admit a
stationary solution at $\log a\to\infty$, because all inflationary
modes reenter the horizon in the asymptotic future, and there is no
entropy associated with inaccessible information (even if we prescribe 
entropy to a decaying de Sitter spacetime, complete recovery of all information encoded in inflationary modes
implies zero final entropy thus violating the second law of thermodynamics).
Note however that if a decelerating FRW expansion stage is followed up by another quasi-de Sitter stage
with a Hubble rate $H_{\Lambda}<H_0$, a fraction of inflationary modes
with $k<aH_{\Lambda}$ will never reenter the horizon, and there will
be an associated remaining entropy $\sim H_{\Lambda}^{-2}$ much higher than
the entropy $\sim H_{0}^{-2}$ of the quasi-de Sitter spacetime realized
during primordial inflation. Ignoring the matching FRW stage of decelerating
expansion, one can simply describe behavior of the universe following
the the dynamics of the effective inflaton field between the minima
of its potential $V(\Phi)$. In the limit $\log a\to\infty$
the lowest among such minima will be the ones dominating the asymptotic
form of the distribution function $P(\Phi)$. Does not it look
like the Universe we are currently living in? \citep{Riess1998,Perlmutter1999,Komatsu2011,Ade2016}
Of course, it depends on the particular form of the effective equation
of state for dark energy whether the current quasi-de Sitter stage
also comes to the end \citep{Copeland2006,Podolsky2002}.

4) \textbf{Do gravitational vector/tensor modes contribute to the
gravitational entropy? }We have just seen that the scalar gravitational
degrees of freedom produce de Sitter entropy with a correct numerical
prefactor $1/4$, and thus the contribution of gravitational tensor
and vector modes to this entropy amounts to zero. The physical reason
for this conclusion is probably that tensors and vectors are unable
to produce a non-trivial large scale superhorizon field structure.
We believe the answer to be naturally the same for a Schwartzschild
black hole, although only deriving and using an analogue of stochastic
formalism for a Schwartzchild spacetime would allow to set the record
straight, which is a task worth investing some time into. We expect the answer to change for an electrically
charged or a rotating black hole. 

5) \textbf{Arrow of time.} Personally, the most physically interesting
consequence of the arguments presented above is related to the nature
of the arrow of time. An entire matter content of our Hubble patch
has the entropy $\sim{}10^{104}k$, while the gravitational entropy
associated with $dS$ horizon (corresponding the present regime of accelerated
expansion) is about ${}10^{122}k$. In the early Universe, the gap between
the gravitational and matter content entropies was even worse, and
we would like to conclude that the analysis of the emergence of the
arrow of time should be performed within the context of quantum gravity
(at least, in the WKB approximation) - inflation does explain why
the matter entropy was so low at Big Bang, but does not explain why
the gravitational entropy was also extremely low at the beginning of inflation.
On the other hand, a huge $dS$ entropy is itself generated only if
the expanding branch of the general solution of the Wheeler-deWitt
equation is picked, as we can see from (\ref{eq:FokkerPlanck}). So,
something else than matter decoherence/gravitational decoherence due
to scattering of gravitational degrees of freedom against matter ones
should be responsible for the arrow of time \citep{Podolskiy2016}.

6) \textbf{IR vacuum restructuring in different theories.} In principle,
the situation with IR vacuum restructuring is not uncommon in physics:
consider even an extremely well studied 4-dimensional quantum electrodynamics, which is famously plagued
with the IR catastrophe problem \citep{Berestetskii2008}. It
is well-known that the latter can be resolved once one takes the bath
of infrared photons into account; photons with very large wavelengths
cannot be practically detected by any physical detector with a finite
bandwidth, and the bandwidth cutoff also serves as an effective IR
cutoff $\Lambda_{{\rm IR}}$ for the theory. In a sense, the true
vacuum state of QED is a formal Fock vacuum state plus the bath
of deep infrared photons. It is interesting to check if there is an
entropy associated with this vacuum degeneracy in QED (naively, there
is since the infrared cutoff in the theory implies tracing out all
degrees of freedom with $\lambda>\Lambda_{{\rm IR}}^{-1}$ and an
associated entropy $\sim\Lambda_{{\rm IR}}^{-2}$ \citep{Srednicki1993}).
In both cases of interest for us (black holes and de Sitter space)
the physical situation is largely similar: there exist modes with
sufficiently large IR wavelengths such that an observer performing
observations of the physical state during a finite interval of time
is unable to descriminate between the formal vacuum state of the field
theory and the state containing an arbitrary ensemble of such IR modes.
For dS spacetime, the modes are the superhorizon modes. In the case
of a spacetime containing an evaporating black hole, such deep infrared
(as perceived by observer at asymptotic spatial infinity) modes are
the ones associated with black hole soft hair.

\end{document}